# Coupling liquid electrochemical TEM and mass-spectrometry to investigate electrochemical reactions occurring in a Na-ion battery anode

Kevyn Gallegos Moncayo[1,2], Nicolas Folastre[1,2], Milan Toledo[1], Hélène Tonnoir[1,2], François Rabuel[1,2], Grégory Gachot[1,2], Da Huo[1,2], Arnaud Demortière[1,2,3*]

[1]Laboratoire de Réactivité et Chimie des Solides (LRCS),
CNRS UMR 7314, UPJV, Hub de l'Energie, 15 rue Baudelocque, 80039 Amiens Cedex, France.
[2]Réseau sur le Stockage Electrochimique de l'Energie (RS2E),
CNRS FR 3459, Hub de l'Energie, 15 Rue Baudelocque, 80039 Amiens Cedex, France.
[3]ALISTORE-European Research Institute,
CNRS FR 3104, Hub de l'Energie, Rue Baudelocque, 80039 Amiens Cedex, France.

*Corresponding author: **arnaud.demortiere@cnrs.fr**

**Abstract:** In this study, we propose a novel approach for investigating the formation of solid electrolyte interphase (SEI) in Na-ion batteries (NIB) through the coupling of *in situ* liquid electrochemical transmission electron microscopy (ec-TEM) and gas-chromatography mass-spectrometry (GC/MS). To optimize this coupling, we conducted experiments on the sodiation of hard carbon materials (HC) using two different setups: *in situ* ec-TEM holder (operating in an "anode free" configuration, referred to as µ-battery) and ex-situ setup (Swagelok battery configuration). In the in situ TEM experiments, we intentionally degraded the electrolyte (NP30) using cyclic voltammetry (CV) and analyzed the recovered liquid product using GC/MS, while the solid product (µ-chip) was analyzed using TEM techniques in a post-mortem analysis. The ex-situ experiments served as a reference to observe and detect the insertion of Na$^+$ ions in the HC, SEI size (389 nm), SEI composition (P, Na, F, and O), and Na plating. Furthermore, the TEM analysis revealed a cyclability limitation in our in situ TEM system. This issue appears to be caused by the deposition of Na in the form of a "foam" structure, resulting from the gas release during the reaction of Na with DMC/EC electrolyte. The foam structure, subsequently transforms into a second SEI, is electrochemically inactive and reduce the cyclability of the battery. Overall, our results demonstrate the powerful synergy achieved by coupling in situ ec-TEM and GC/MS techniques, which provides a deeper understanding of the dynamics and behavior of SEI. Consequently, this knowledge contributes to the advancement of the new generation of NIB. Moreover, the impact studies of new electrolyte molecules on battery operation and safety, as well as the evaluation of premature degradation, are fundamental aspects for the development of a commercially viable and eco-compatible energy storage technology.

**Introduction**

The growing demand for energy storage devices has become a key issue for our well-being. Among the solutions, Li-ion batteries (LIBs) are widely used in portable consumer electronics and electric vehicles due to their high-density energy, high cycle durability, and fast speed of charge. However, the increasing demand for LIB has led to rising prices and increased pressure on Lithium[1]. To find a cleaner and more sustainable alternative, interest has again been focused on Na-ion batteries (NIBs). NIB technology was already explored in the 80s, which was abandoned due to the better performance of LIB[2,3,4].

NIB is now considered as a promising technology because Na is naturally abundant, cheap, and environmentally friendly. NIB batteries allow the use of Al foil as a current collector in both anode and cathode, which, in comparison with LIB (use of Cu current collector at low voltages) represents a reduction in costs of fabrication[5]. In addition, NIBs offer high power output and fast charging for applications such as electric bicycles, scooters, mopeds, robots, industrial tools. NIB batteries also can be considered as an alternative to LIB in terms of long-term energy storage[6]. But they are not competitive in terms of electrical energy density[7].

Thus, in recent years, to improve their energy performance, different methods have been used to characterize them during the charge/discharge process, such as XRD, XPS, NMR, SEM and, TEM[2,8,9]. However, the dynamic conditions within a battery had not been explored due to the requirement of a sealed working environment, until the advent of the *in-situ* liquid ec-TEM (electrochemical-TEM) characterization technique[10,11,12]. This is a powerful tool for real-time liquid TEM imaging applications that provides a deeper understanding of the behavior of materials in the electrochemical liquid environment, such as oxidation-reduction reactions, nucleation, crystalline phase transformations, the kinetics of chemical species formation and, movement of nanometric objects in the fluid[2,13,14].

The study of the dynamics of electrochemical regimes of NIBs with high spatial resolution at the nanoscale is fundamental to further understanding these energy storage systems. For example, Lutz L. *et al.* have evidenced, thanks to the *in situ* liquid ec-TEM technique, the development of parasitic reactions in Na-O$_2$ batteries, which could be responsible for the low charge capacity and poor cycling stability of the batteries[12]. On the other hand, to reinforce this *in situ* liquid ec-TEM technique, we propose a coupling with the mass-spectrometry to analyze the products formed during the charge and discharge of these batteries. Studies carried out with GC/MS by G. Gachot *et al.* detected and accurately identified a wide range of volatile molecules resulting from battery degradation, demonstrating that CH$_3$OLi is the initiator of electrolyte degradation in lithium-ion batteries[15,16,17].

By using the *in situ* liquid ec-TEM and the GC/MS coupling, we intend here to optimize this coupling by investigating the electrochemical behavior of the hard carbon (HC) electrode in contact with the NP30 electrolyte[18]. Thus, to improve the performance of the NIB, it is necessary to find optimal compromises between electrolyte compositions and electrode surface states, since the formation of the protective layer of SEI (Solid Electrolyte Interphase) depends on them. The selection of appropriate electrolytes is particularly crucial as it governs the composition of the SEI, influencing its properties including ionic conductivity, electronic resistance, stability, and thickness.[19,20]

However, even whether the electrolyte formulation is studied as a crucial factor for the quality of the life cycle, the SEI is not well known for the NIB. Previous studies[21] indicate that secondary reactions as the degradation of electrolytes take place in the first cycle of the battery, which leads to the formation of SEI on the hard carbon surface. From a thermodynamic point of view, the cathodic stability of an electrolyte depends on its level of lowest unoccupied molecular orbital (LUMO). The electrolyte component with the lowest LUMO level will reduce first in contact with the HC electrode, possibly contributing to SEI formation[21].

Several authors have studied the degradation and formation mechanisms of SEI and electrolyte degradation in LIB systems by using *in-situ* liquid ec-TEM and GC/MS analysis separately. For instance, Robert L. Sacci *et al.*[22] studied the SEI formation using *in-situ* liquid ec-TEM however without reporting analysis on the degradation of the electrolyte. In addition, other authors were interested in the comprehension of electrolyte degradation and used both, TEM and GC/MS techniques, but exclusively *ex-situ* in coin cell batteries and without synergy between TEM and GC/MS[23,24,25]. This study presents the novel investigation of SEI formation and electrolyte degradation during Na-ion insertion into hard carbon materials in a

NIB system. To our knowledge, this is the first report utilizing a simultaneous approach involving ec-TEM and GC/MS techniques. The synergistic combination of these techniques offers a promising methodology for gaining deeper insights into the underlying phenomena of NIBs. Furthermore, this approach can be extended to optimize other energy storage materials, thus expanding its potential as a valuable tool in the field of energy storage research.

**Experimental Methodology**

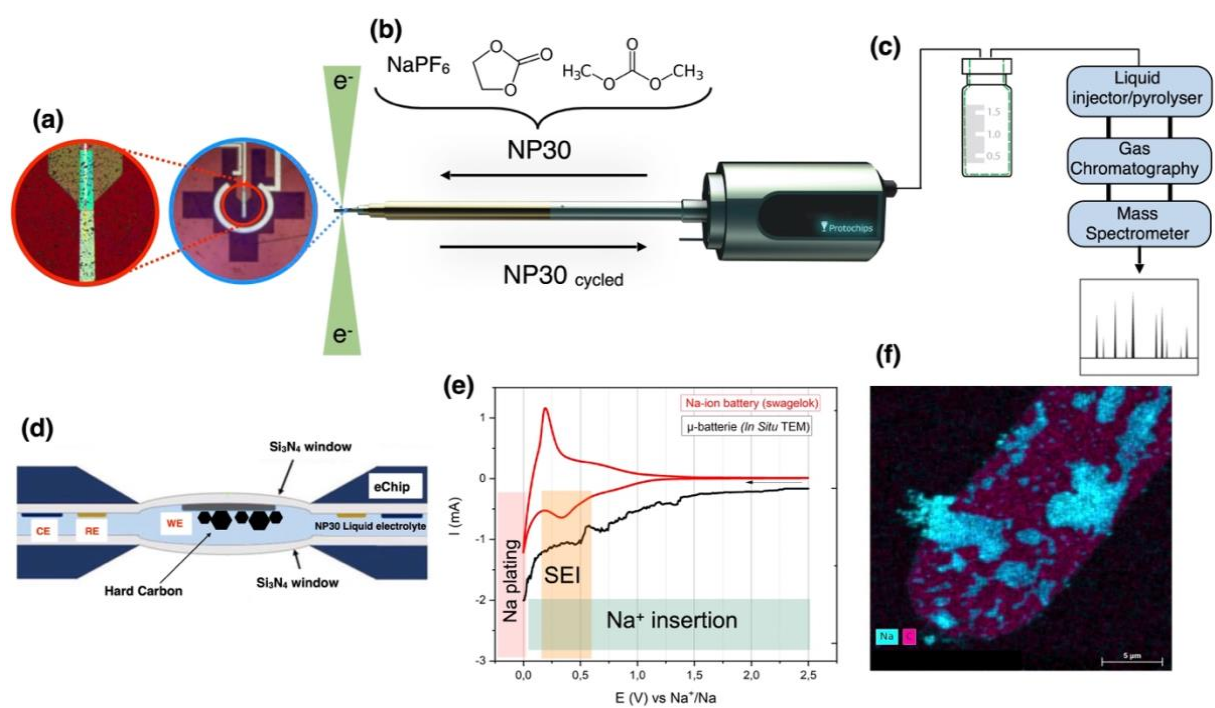

**Figure 1**: Methodology process for liquid electrolyte degradation analysis. (a) real image of electrodes in a microchip for electrochemical measures, working electrode (WE), reference electrode (RE), and counter electrode (CE). (b) Protochips sample holder for liquid ec-TEM image *in situ* analysis. (c) Hermetic flask for the cycled liquid electrolyte for further analysis and expected chromatograph after analysis of volatile liquid electrolyte in GC/MS. (d) Schematic representation of *in situ* electrochemical liquid cell with Hard Carbon forming µ-battery. (e) Comparison of cyclic voltammetry of Na-ion battery vs µ-battery, including $Na^+$ insertion, SEI formation and Na plating. (f) Results from EDX-STEM imaging in liquid TEM cell using µ-battery.

For a better understanding of electrolyte degradation, an adapted configuration was conceived to perform close-to-reality battery cycling, as shown in Figure 1. The liquid electrolyte NP30 (hexafluorophosphate $NaPF_6$ 1M, in a mix of ethylene carbonate and dimethyl carbonate EC: DMC 1: 1 in mass ratio) was injected via micro fluid tubes to the ec-TEM cell for cycling in an HC/Na half cell, as shown in Figure 1b and Figure 1d. The electrochemical measurements are performed with a 3-electrode system displayed in Figure 1a, and the results are presented in Figure 1e as a cyclic voltammetry (CV) scanning from 0.0

V to 2.5 V (vs. Na+/Na), with a sweeping rate of 1 mV/s. Immediately after cycling and while the cell is still inside the TEM chamber, TEM imaging was performed, and the cycled electrolyte was recovered after the first discharge in a hermetic flask for further analysis in GC/MS, as shown in Figure 1c. This process reduces the perturbation in each step of data acquisition, especially the further degradation of the electrolyte due to environmental exposure, and reflects as well a less biased result and in consequence, allows a better interpretation of the phenomenon observed.

### Half-coin cell battery

Cyclic voltammetry was performed in a half coin cell (HC/Na metal) to obtain a global view of the electrochemical behavior of the configuration anode/electrolyte used here as an experimental reference for the *in situ* investigation. The counter-electrode was metallic Na. The electrolyte used for the coin cell was NP30 (NaPF$_6$ 1M, EC: DMC 1: 1 mass). Electrochemical analysis was performed using SP200 potentiostat. EC-lab software was used for controlling the device. The scan conditions are those presented before and the conditions are identical for all the electrochemical curves performed during this study.

### EC-TEM Electrode preparation

The active material (commercial HC suspended in methanol) is deposited on the top e-chip through the drop-casting method (20-200 µl taken with a micropipette and deposited over the µ-chip and left to dry for 24 hours at room temperature).

### µ-Battery Assembly

The head of the *in-situ* liquid ec-TEM sample holder contains a µ-fluidic tubing system and an electrical wiring system for electrochemistry, as shown in Figure 1d. The electrochemical liquid cell, also called µ-battery in this work, is prepared from the assembly of two seals: a bottom e-chips and a top e-chip, which are in contact of O-ring polymer to garanty a good sealing. There are 3 coplanar electrodes on the top e-chip: a WE (glassy carbon), RE (Pt), and CE (Pt). The HC active material is deposited on the glassy carbon surface the WE electrode of the µ-chip. This assembly is held in place by a metal hood that is adjusted with a torque screwdriver.

### TEM analysis

TEM and STEM analyses were performed by the TECNAI F20-S-TWIN, FEI device, and parameter were adjusted as follow: 200 KV for the acceleration tension, OneView camera

(Gatan), spot size 5 and a 70 μm condenser aperture. Image techniques applied over the sample were TEM image, STEM-EDX (Scanning Transmission Electron Microscope Energy Dispersion X-ray), and STEM-HAADF (High Annular angle Dark Field). The data acquired was treated using Digital micrograph software (Gatan) and FIJI software.

Regarding the μ-chip battery (*in-situ*), images were obtained at the end of the cyclic voltametry. After imaging, the μ-battery cell and the μ-fluid tubes were cleaned using DMC, then methanol. The drying procedure was performed during the night in vacuum conditions.

For *ex situ* analysis, a Swagelok-type battery (HC/Na) was assembled and studied in the same conditions as for *in situ* TEM imaging. After one cycle, the HC electrode was removed from the Swagelok cell in the dry room and rinsed several times with DMC. The images were acquired after the battery cycling process (post-mortem) to compare those results to the *in situ* analysis.

*GC/MS*

The acquisition parameters in GC / MS (Thermo Scientific): For GC, a flow rate of 1.5 mL/min. of He and a temperature gradient[15]. For MS, an electronic impact (EI) source with an ionization energy of 70 eV. Use of a quadrupole (Q) to separate compounds based on their mass/charge ratio (m/z) (identification of compounds being processed with the National Institutes of Standards Library (NIST)).

**Results and Discussion**

**Study of the electrochemical system on the anode side (HC) - Half coin cell**

The analysis over half coin cell was performed to have a better understanding of the system before performing the analysis in the μ-battery. As shown in Figure 2a, cyclic voltammetry allows us to observe different electrochemical reactions represented in three zones over two cycles, corresponding to the (dis)insertion of Na+ in the HC anode (green), the formation of the SEI (orange), and the Na plating on the anode surface (red). As the potential decreases from 2.5 V (Vs. Na$^+$/Na) down to the anodic peak (A) at 0.2 V, the positive current increases significantly while Na$^+$ disinsertion from HC. Similarly, Na$^+$ insertion takes place in the (C1) region while negative current increases toward high potential.

During the first discharge cycle, a cathodic peak (C2) is observed *at* 0.3 V, corresponding to the formation of the SEI. However, during the second discharge cycle, the cathodic peak (C2) disappears as the reduction of the electrolyte (NP30) during the formation of the SEI in the first discharge cycle is so far irreversible.

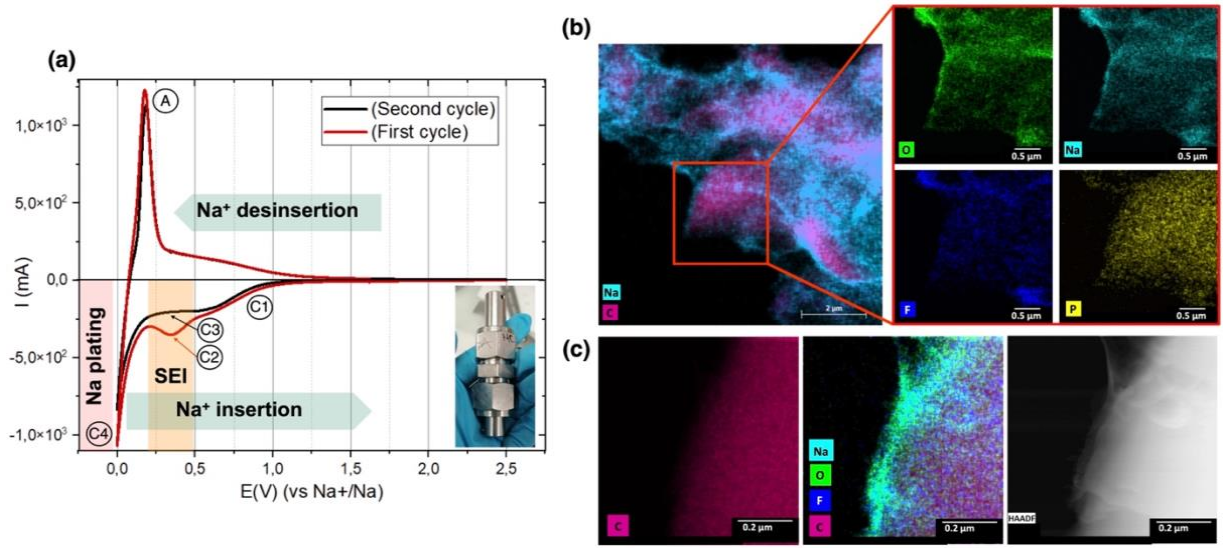

**Figure 2** : (a) Cyclic voltammetry (voltage vs current) of a half coin cell at 1 mv/s with (A) anodic peak after disinsertion of Na+, (C1) Insertion of Na+ in HC, (C2) Formation of SEI, (C3) disappearing of the SEI formation cathodic pic during second discharge, and (C4) Na plating. (b) STEM-EDX of the hard-carbon anode after the second cycle showing Na plating. (c) STEM-EDX and HAADF of the SEI on the HC surface.

Moreover, in the work of L. Zhang et al[26]. it was determined that the interaction between the electrolyte in metallic Na leads to the formation of gas in an open circuit. As a result, the coin cell battery presents two kinds of electrolyte degradation: Na/NP30 degradation and SEI formation, which explains the low cycling life in NIBs.[27,28,29]

As the potential gets closer to the standard potential of $Na^+$/Na, the high current density on the anode surface and the low ionic conductivity of the HC for $Na^+$ ions lead as expected to Na plating on the HC particles surface. This pronounced reduction in current corresponds to the (C4) peak in Figure 2a.

*In-situ analysis - µ-battery electrochemical behavior*

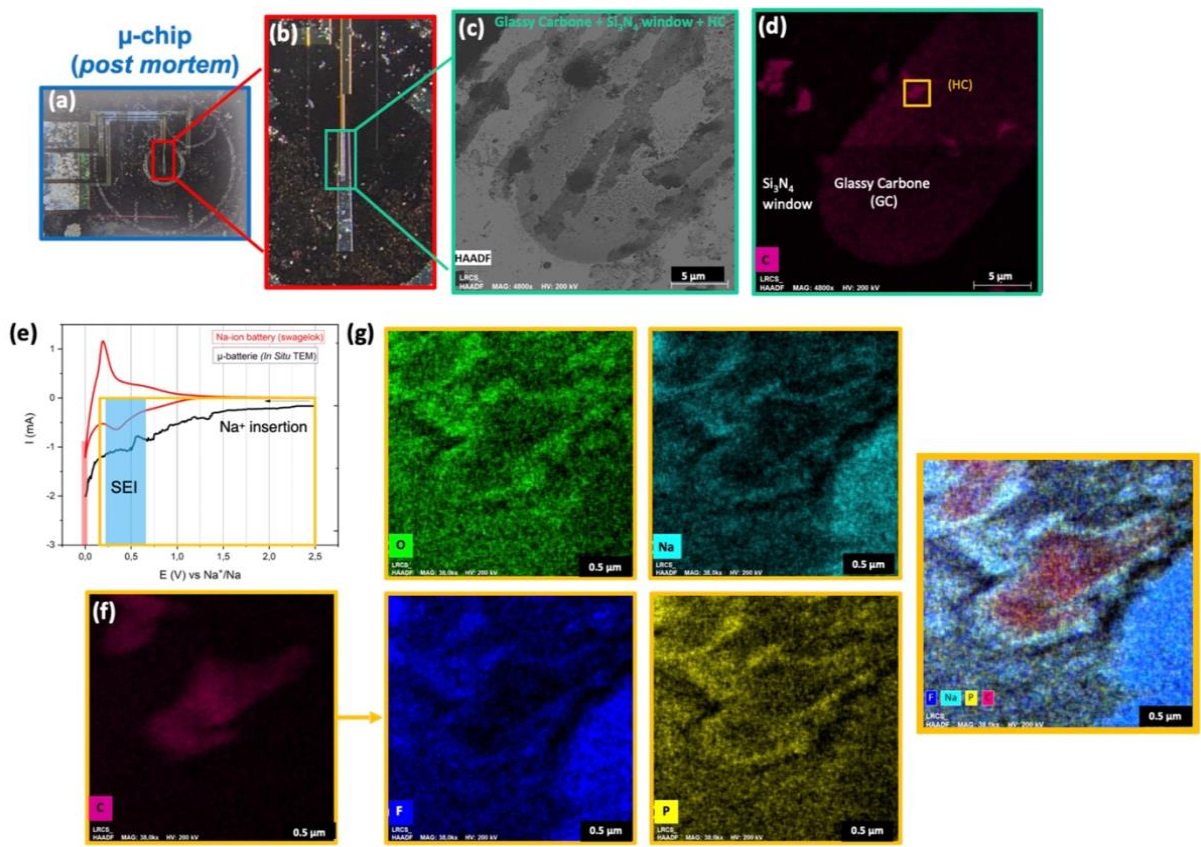

*Figure 3: (a) µ-chip post-mortem image obtained by optical microscope, (b) Zoom in over WE, (c) HAADF image over the WE, (d) EDX cartography over WE to reveal the HC particles, (e) Voltamperogram of coin cell and µ battery: Zone 1 (yellow): Continuous insertion of Na+ into the HC. Zone 2 (blue): Formation of SEI. Zone 3 (red): Plating deposit of Na, (f) EDX zoomed-in image over an HC particle, (g) EDX image with detection of different elements (O, Na, F, P).*

The µ-battery has been opened post-mortem to characterize the HC anode after cycling, as shown in the optical microscope images in Figure 3a and Figure 3b. The elements of the WE such as the glassy carbon, $Si_3N_3$ window, and the earlier deposited HC anode can be distinguished in the HAADF and STEM-EDX, as presented in Figure 3c and Figure 3d respectively. After cycling, the EDX elemental cartography has been performed on isolated HC particles on the glassy carbon WE to better understand the degradation/formation mechanisms. Thus, the analysis of SEI formation is presented in Figure 3f on one of the localized HC particles in Figure 3d region. The EDX map reveals a layer surrounding the particle, composed of Na, O, F, and P and that presents a thickness of 389 nm. In addition, the voltammogram curves of the coin cell and the µ-battery follow similar trends, despite the instability of the current and the background noise. This background noise may have multiple sources such as the electrolyte flow rate and the reference stability, as the electrochemical measurements were performed with a flow rate of NP30 of 1 µL/min and a sweeping rate of 1 mV/s with a Pt pseudo-reference.

EDX imaging reveals a higher concentration of Na around HC particles in comparison with the bulk of the HC particle as well. In the work of Z. Wei *et al.*, it was found that SEI is mostly composed of organic and inorganic compounds, being the organic compounds (sodium alkyl carbonates; $RCH_2OCO_2Na$) and inorganic products ($Na_2O$, $NaF$), both Na-rich compounds.[30] Thus, the strong presence of Na around the HC anode particles may result from the SEI formation.

Several hypotheses can be formulated: 1) Due to the lack of Na in the µ-battery configuration, degradation of NP30 is not present at open circuit voltage, meaning that the electrolyte is degraded by SEI formation only during the cycling process. 2) Close to 0 V (vs $Na^+/Na$), the sudden drop in current represents an interaction between $e^-$ and $Na^+$ for plating layer formation, which then reacts with the electrolyte to form a secondary SEI ($SEI_{bis}$). The gas released by the interaction leads to the formation of a foam-like structure, as shown further in Figure 5f.

Moreover, in Figure 3g, Na is found inside the HC particles. This observation points to the insertion of Na during cycling, in well-agreement with the electrochemical curve. $Na^+$ insertion in HC follows different mechanisms such as sodium ion insertion on structural defects, $Na^+$ ion intercalation between graphene layers, and micropore insertion, as presented in the work of H. Tonnoir et al[31].

***Electrolyte degradation analysis***

After studying the electrochemical systems of HC in the µ-battery, the liquid product was recovered for analysis by GC/MS.

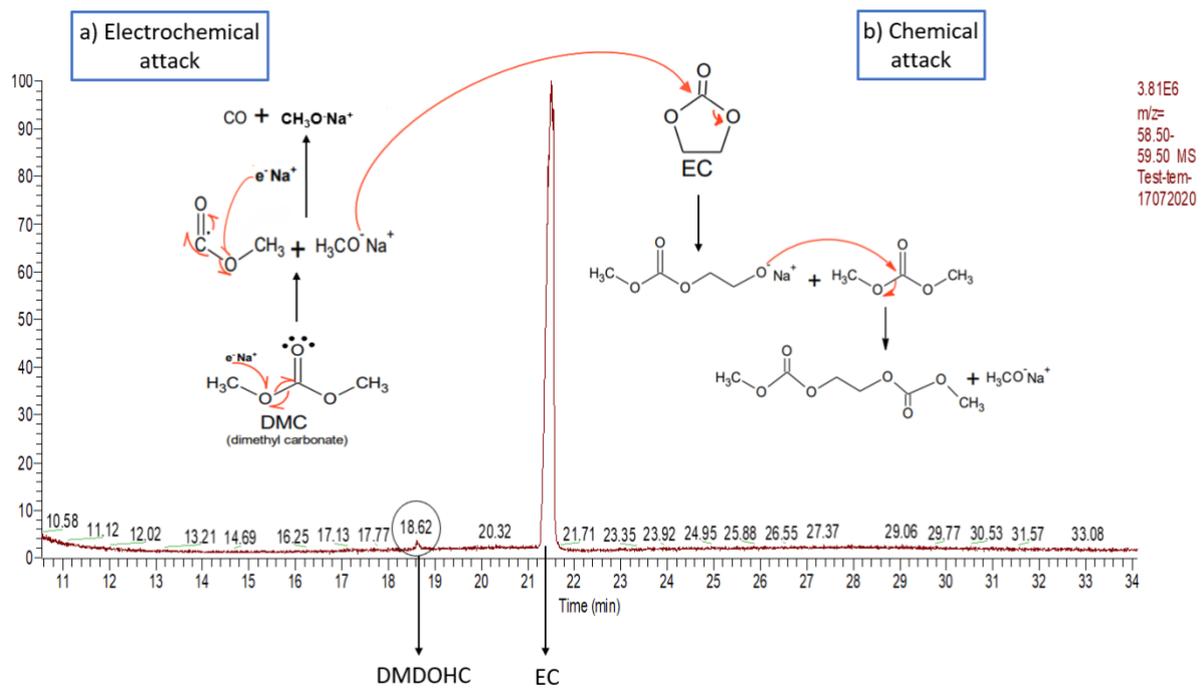

*Figure 4: Chromatogram obtained for the liquid electrolyte after degradation. Specific mass m/z = 59, calibration for detection of DMDOHC (dimethyl-2.5-dioxahexane carboxylate plus sodium methoxide). (a) electrochemical attack carried by the e⁻ of Na atoms. (b) chemical attack carried by the interaction of $H_3CONa$ (salt).*

Figure 4 presents the chromatograph obtained in which at 18.62 min DMDOHC is observed. The results are coherent with the literature.[32,33,34] DMDOHC is a common degradation product in commercial batteries that use LP30 as electrolyte: ($LiPF_6$ at 1 mol/L in a mix of EC:DMC = 1:1 mass ratio).

The reaction mechanism between DMC/EC and Na is presented in Figure 4 as well. This is one of the possible paths that can lead to the formation of the degradation product. However, several different mechanisms can take place at the same time in parallel[15]. The first step is an electrochemical reduction due to a one-electron attack produced by Na e⁻ over an oxygen atom in DMC due to the electronegativity of O, this results in sodium methoxide formation ($H_3CONa$) plus a free radical ($C_2O_2H_3$). A second electrochemical attack takes place between the free radical and Na e⁻ giving as a result the liberation of CO plus sodium methoxide. A chemical reaction takes place as well between EC and sodium methoxide. The O in the methoxide attacks the C surrounded by 3 O, opening the molecule. The product of this chemical reaction reacts with DMC to obtain dimethyl-2,5-dioxahexane carboxylate plus sodium methoxide[35].

As presented in the mechanism, DMDOHC is the result of the interaction between metallic Na and solvents in NP30, which means that during cyclin Na metal is formed and reacts immediately with the electrolyte. However, according to the electrochemical curve, metallic Na could be formed only close to 0 V (vs Na$^+$/Na) (plating). A possible explanation is that during SEI formation, metallic Na exists, without a plating process, and reacts with solvents, meaning that SEI formation leads to DMDOHC formation, however, this hypothesis has to be proven by further analysis.

*Na plating*

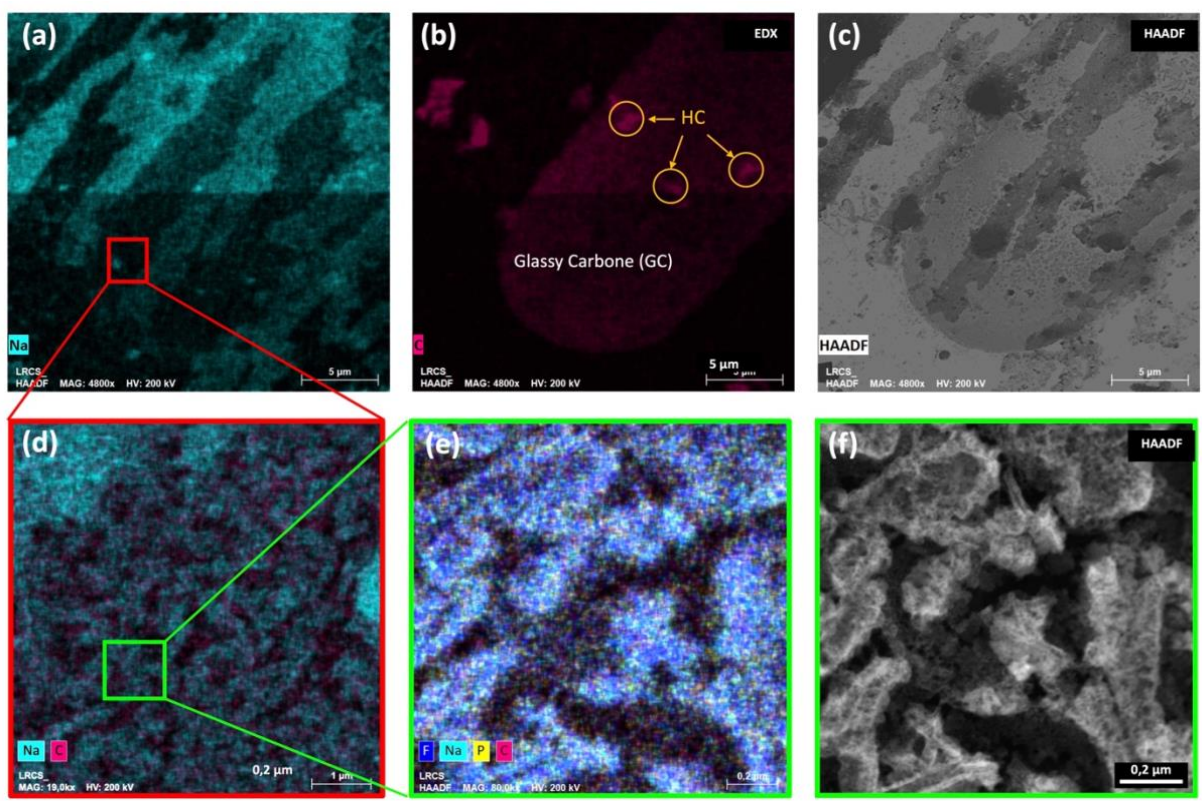

*Figure 5*: EDX maps of (a) Na and (b) Carbon, and (c) HAADF analysis over the same area of the WE. (d) Zoom in a Na-poor zone which reveals a dendritic kind of structure, (e) EDX analysis at higher magnification of image d where a dendritic kind of structure is observed, (f) HAADF over the same zone as image d where foam type structure is founded.

As shown in Figure 5, an uneven distribution of Na is observed over the WE. The EDX analysis over the edge of the electrode reveals a dendritic-like structure with a low concentration of Na, as shown in Figure 5d. The morphology can be better appreciated in the magnified HAADF image revealing the formation of a foam-like structure in the dendrites, as shown in Figure 5f. In addition, the EDX analysis of the structure shows the presence of the same elements as the SEI layer.

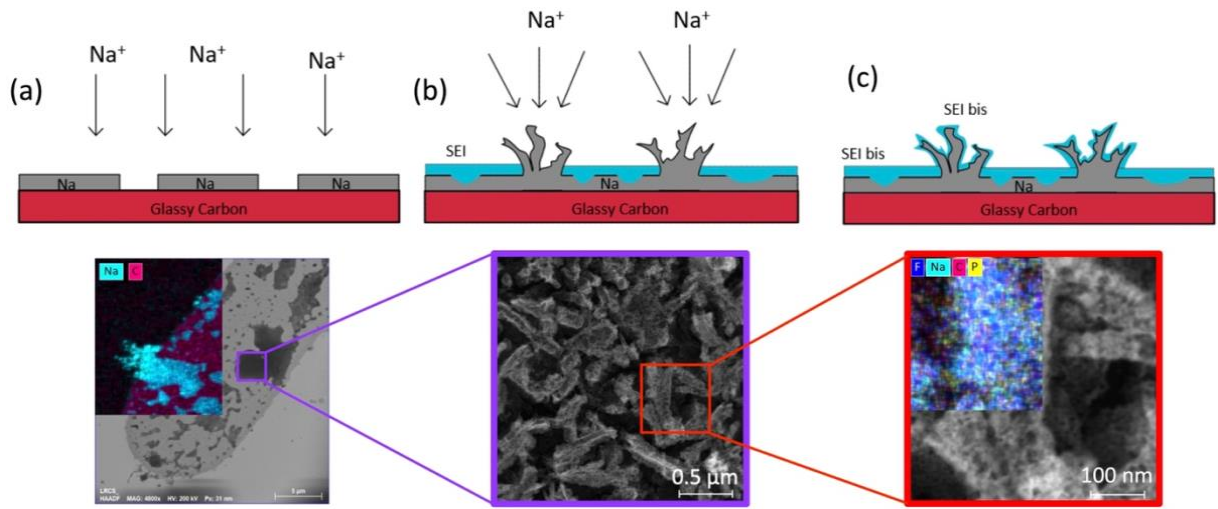

*Figure 6*: Schematic representation of electrochemical reaction occurring on the surface of glassy carbon electrode with STEM images associated. reactions (a) non-uniform Na plating, (b) dendrite and SEI formations, and (c) SEI$_{bis}$ formation.

During Na plating, a non-uniform layer of metallic Na is deposed over the glassy carbon WE, as shown in Figure 6a. The formation of this layer creates different nucleation points which leads to the formation of dendrites shown in Figure 6b. Then, the metal reacts with the electrolyte and forms the SEI$_{bis}$ mentioned previously, as shown in Figure 6c. The presence of metallic Na as well as the formation of the SEI$_{bis}$ liberates gases that contribute to the formation of the foam-like structure. The SEI$_{bis}$ is formed over the dendrites that are the substrate for the protective layer and adopt its form. Even though the SEI$_{bis}$ is conductive for ions, the foam structure reduces its conductivity and as a consequence the performance of the battery. In this case, during plating, the two kinds of electrolyte degradation processes of the half-coin cell are present.

## Ex-situ analysis - Swagelok electrochemical behavior

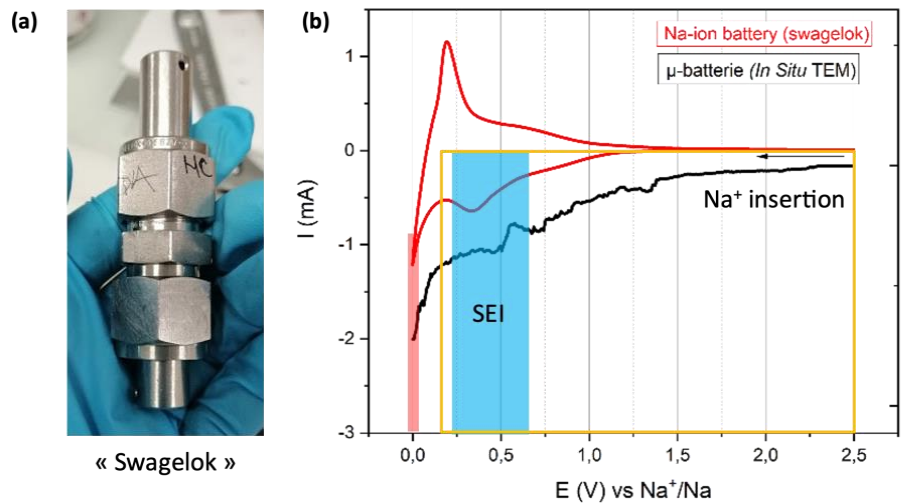

*Figure 7: (a) Swagelok cell configuration for cycling and post-mortem analysis, (b) Electrochemical curve for the Swagelok and the µ-battery experiments (voltage vs current). The shape of the discharge curve is similar for both configurations.*

As shown in Figure 7a, electrochemical analysis was carried out in a Swagelok-type battery after one cycle (post-mortem) to compare the results with the *in-situ* analysis. The cycling was performed with the same parameters as for the µ-chip. The voltammogram presented in Figure 7b shows similar trends for Swagelok, coin cell, and µ-battery curves. These similarities in the electrochemical behavior allow us to compare the *ex-situ* and *in-situ* results for validation. In addition, the counter electrode for Swagelok was metallic Na, meaning that the two kinds of electrolyte degradation present in the half-coin cell configuration may be present as well.

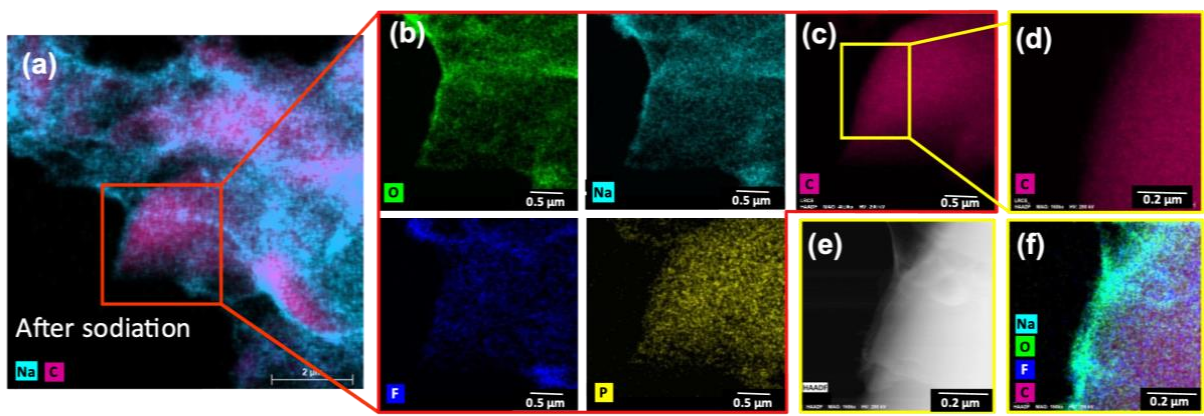

*Figure 8: (a) TEM-EDX analysis over a particle obtained after cycling in Swagelok. (b) EDX zoomed-in image over an HC particle to represent the different species present in the particle and its surroundings, (c) HC particle (EDX view) to highlight the Carbon*

*domain, (d) EDX image over the edge of HC particle, (e) HAADF TEM image of the SEI layer and HC edge for the determination of SEI thickness,( f) EDX image of the SEI layer surrounding the HC particle.*

| Element | Mass ratio | Atomic ratio |
|---|---|---|
| Oxygen | 33.36 % | 49.93 % |
| Sodium | 43.16 % | 35.04 % |
| Phosphate | 04.61 % | 08.18 % |
| Fluorine | 10.33 % | 06.84 % |

**Table 1:** *Semi-quantitative values for TEM-EDX analysis over SEI post-mortem Swagelok battery configuration. The SEI represents 11.8 % of the total imaged surface of the HC particle.*

As shown in Figure 8 and Table 1, two major species Na and C are detected over an HC particle, according to the EDX analysis. Furthermore, the detection of Na inside the C particle confirms the insertion of Na in HC. A zoom-in over the HC particles reveals a layer formed of various species, as shown in Figure 8f. This result is coherent with the EDX of the *in-situ* analysis, confirming the formation of an SEI layer surrounding the HC anode particle, as the layer is composed of the same elements found in the SEI for the *in-situ* analysis. However, the thickness of the SEI is only 83 nm in the Swagelok battery, a considerable reduction in comparison with the µ-battery case. The lower thickness of the SEI in the Swagelok configuration can be explained by the difference in current distribution during cycling. Indeed, the necessary pressure applied in the Swagelok to enhance the contacts between the different battery components induces a significant variation in the current distribution in comparison with the µ-battery configurations. A zoom-in on the EDX cartography of the HC particle shows the chemical contrast between the HC and the SEI, as shown in Figure 8f.

**Conclusion**

Na-ion batteries (NIB) exhibit great promise for future applications in the field of energy storage due to the abundance and low cost of their base materials, as well as their environmentally friendly properties. However, in order to progress towards commercialization, it is crucial to delve deeper into the limitations of NIB.

In this study, we employed the coupling of liquid electrochemical transmission electron microscopy (ec-TEM) and gas chromatography/mass spectrometry (GC/MS) to gain a comprehensive understanding of solid electrolyte interphase (SEI) formation and the degradation of liquid electrolyte NP30. This combined approach served as a unidirectional analysis method, enabling us to track the battery's cycling behavior and enhance our comprehension of the chemical and electrochemical phenomena involved.

To summarize, we utilized a liquid electrochemical cell for in-situ µ-battery measurements, a coin cell battery with an HC anode as an electrochemical reference, and a Swagelok battery for ex-situ measurements to validate the in-situ analysis. Electrochemical analysis revealed three distinct stages of behavior during the initial cycle: Na+ insertion, SEI formation, and Na plating near 0 V. The cyclic voltammetry data demonstrated the coherence of the *in situ* and *ex situ* experiments with the coin cell reference, showcasing similar electrochemical behaviors. Additionally, *in situ* energy-dispersive X-ray spectroscopy (EDX) maps confirmed the formation of the SEI layer around HC particles, aligning with the electrochemical curves. Moreover, gas chromatography/mass spectrometry (GC/MS) analysis detected the presence of DMDOHC, a product resulting from NP30 electrolyte degradation, which was consistent with previous findings in the literature. Our findings indicated that the formation mechanism of DMDOHC from electrolyte degradation leads to the generation of gases such as CO, contributing to the foam-like morphology of Na plating as observed through TEM-EDX and confirmed by *ex situ* results. Furthermore, the formation of SEI through this process further reduced the battery's performance.

The successful coupling of ec-TEM and GC/MS proved advantageous, as it minimized disturbances in imaging and electrolyte analysis. This analytical synergy holds great potential for future research endeavors, offering an improved understanding of SEI formation, electrolyte degradation, and the formulation of new NIB configurations to enhance battery performance.

Further investigations are warranted to deepen our understanding of SEI formation and electrolyte degradation in NIB. Specifically, the analysis of gases should be conducted to confirm the presence of the predicted degradation products, and in-situ analysis at different states of charge (SoC) should be performed to gain insights into the dynamics of SEI formation. Moreover, employing the same methodology and configuration with different electrolytes will facilitate the optimization of the anode/electrolyte pairing for NIB.

**ACKNOWLEDGMENT**

The authors thank the French National Research Agency (ANR) for their financial support in the project DESTINa-ion_Operando project ANR-19-CE42-0014-02. Additionally, the UPJV and RS2E electron microscopy platform were utilized for this research. The authors acknowledge funding from the RS2E network via the STORE-EX Labex Project ANR-10-LABX-76-01.

**COMPETING INTERESTS**

The authors declare no competing interests.